\documentclass[conference]{IEEEtran}

\IEEEoverridecommandlockouts
\usepackage{cite}
\usepackage{amsmath,amssymb,amsfonts}
\usepackage{algorithmic}
\usepackage{graphicx}
\usepackage{textcomp}
\usepackage{xcolor}
\usepackage{threeparttable}
\usepackage{amssymb}% http://ctan.org/pkg/amssymb
\usepackage{pifont}% http://ctan.org/pkg/pifont
\usepackage{amssymb}
\usepackage{siunitx}
\usepackage{multicol}
\usepackage{subfig}
\usepackage{booktabs} % To thicken table lines
\usepackage{multirow}
\usepackage{threeparttable}
\usepackage{amssymb}% http://ctan.org/pkg/amssymb
\usepackage{pifont}% http://ctan.org/pkg/pifont
\usepackage{amssymb}
\usepackage{siunitx}
\usepackage{multicol}
\usepackage{subfig}
\usepackage{enumitem}
\usepackage{wasysym}
\usepackage{framed}
\usepackage{pgfgantt,rotating}
\usepackage{subfloat}
\usepackage{multirow}
\usepackage{blindtext}

\usepackage[ruled,%
            linesnumbered]{algorithm2e}
\SetKwInput{KwInput}{Input}
\usepackage{amssymb}

\def\BibTeX{{\rm B\kern-.05em{\sc i\kern-.025em b}\kern-.08em
    T\kern-.1667em\lower.7ex\hbox{E}\kern-.125emX}}
\begin{document}

\title{Evaluation of Encoding Schemes on Ubiquitous Sensor Signal for Spiking Neural Network\\
}
\author{
Sizhen Bian, Elisa Donati and Michele Magno  \\
sizhen.bian@pbl.ee.ethz.ch, elisa@ini.uzh.ch, michele.magno@pbl.ee.ethz.ch
}

\maketitle

\begin{abstract}
Spiking neural networks (SNNs), a brain-inspired computing paradigm, are emerging for their inference performance, particularly in terms of energy efficiency and latency attributed to the plasticity in signal processing. %In ubiquitous scenarios, where the majority of devices are battery-powered, for instance, in wearable systems and haptic interfaces, real-time feedback and power-saving computing are crucial performance factors.  %According to existing findings, SNNs have great potential to improve overall computing efficiency and latency. 
To deploy SNNs in ubiquitous computing systems, signal encoding of sensors is crucial for achieving high accuracy and robustness. Using inertial sensor readings for gym activity recognition as a case study, this work comprehensively evaluates four main encoding schemes and deploys the corresponding SNN on the neuromorphic processor Loihi2 for post-deployment encoding assessment.
% Four main encoding schemes and their variations are explored on a gym activity recognition dataset.
Rate encoding, time-to-first-spike encoding, binary encoding, and delta modulation are evaluated using metrics like average fire rate, signal-to-noise ratio, classification accuracy, robustness, and inference latency and energy. In this case study, the time-to-first-spike encoding required the lowest firing rate (2\%) and achieved a comparative accuracy (89\%), although it was the least robust scheme against error spikes (over 20\% accuracy drop with 0.1 noisy spike rate). Rate encoding with optimal value-to-probability mapping achieved the highest accuracy (91.7\%). Binary encoding provided a balance between information reconstruction and noise resistance. Multi-threshold delta modulation showed the best robustness, with only a 0.7\% accuracy drop at a  0.1 noisy spike rate. This work serves researchers in selecting the best encoding scheme for SNN-based ubiquitous sensor signal processing, tailored to specific performance requirements.
\end{abstract}


\section{Introduction}
\label{sec:intro}

\begin{table*}[t]
\centering
     \begin{threeparttable}

\caption{Related neuromorphic works in ubiquitous sensor data processing with different encoding schemes}
\label{relatedwork}
\small
\begin{tabular}{ p{1.4cm} p{1.8cm} p{1.5cm} p{3.0cm}  p{2.0 cm} p{3.5cm} }
\toprule 
Year-Work & Application  & Sensor  & Encoding Scheme & Training  & Performance Metrics   \\ 
 \midrule 

2019-\cite{corradi2019ecg} & Heartbeat Classification &  ECG & Delta Modulation & SVM+ rSNN  & Accuracy \\
\hline

2019-\cite{blouw2019benchmarking} & Key Word Spotting &  Microphone & (Not Available) & ANN-to-SNN  &  Accuracy, Latency, Energy \\
\hline

2020-\cite{ceolini2020hand} & Hand Gesture Recognition &  EMG & Delta Modulation & Direct SNN  & Accuracy, Latency, Energy \\
\hline

2021-\cite{sharifshazileh2021electronic} & Oscillation Detection &  EEG & Delta Modulation & Direct SNN  & Accuracy\\
\hline

2023-\cite{bos2023sub} & Audio Classification &  Microphone & Amplitude Encoding & Direct SNN & Accuracy, Latency, Energy \\
\hline

2023-\cite{bian2023evaluating} & Activity Recognition &  IMU, Capacitive & Multi-Threshold Delta Modulation & Direct SNN &  Accuracy, Latency, Energy \\

\bottomrule
\end{tabular}
%\begin{tablenotes}
%\setlength{\columnsep}{0.8cm}
%\setlength{\multicolsep}{0cm}
%  \begin{multicols}{2}
%            \item[a] Not Available.
            %\item[b] Sensor power only.
            %\item[c] Depending on the dvs and time window lenght.    
            %\item[d] End-to-end power. 
%  \end{multicols}
%\end{tablenotes}
\end{threeparttable}

\end{table*}

Spiking neuron networks (SNNs)  are gaining popularity for processing two-dimensional event streams from Dynamic Vision Sensor (DVS)~\cite{lichtensteiner2008128x128, mcreynolds2022experimental, bonazzi2024retina}, partly because DVS is the only commercially available event sensor. Prior works have shown SNNs deliver impressive performance on neuromorphioc platforms for one-dimensional signals like electroencephalogram (EEG) and electromyography (EMG) in terms of latency and power consumption~\cite{yang2023neuromorphic, ceolini2020hand, donati2019discrimination, vitale2022neuromorphic}. In ubiquitous computing, the advantage of energy and low latency of neuromorphic somatosensory and computing systems~\cite{donati2024neuromorphic, christensen20222022} are highly beneficial for battery-operated devices with limited hardware resources~\cite{barbuto2023disclosing, bonazzi2023tinytracker, moosmann2023ultra}. Those features are crucial for devices requiring timely reactions~\cite{bhat2023ultralow, dabbous2021touch, bartolozzi2016robots}. 
To process signals in a neuromorphic processor using SNNs, an encoding method is needed to transform output signals into spike streams~\cite{muller2024win}. Moreover, it is crucial to evaluate the performance of different encoding schemes comprehensively and straightforwardly to develop closed-loop neuromorphic ubiquitous systems
~\cite{zai2015reconstruction, rastogi2023spike, chamorro2023event, ferreira2021neuromorphic, liu2010use, rueckauer2018conversion}.

\begin{figure}
    \centering
    \includegraphics[width=1.0\linewidth]{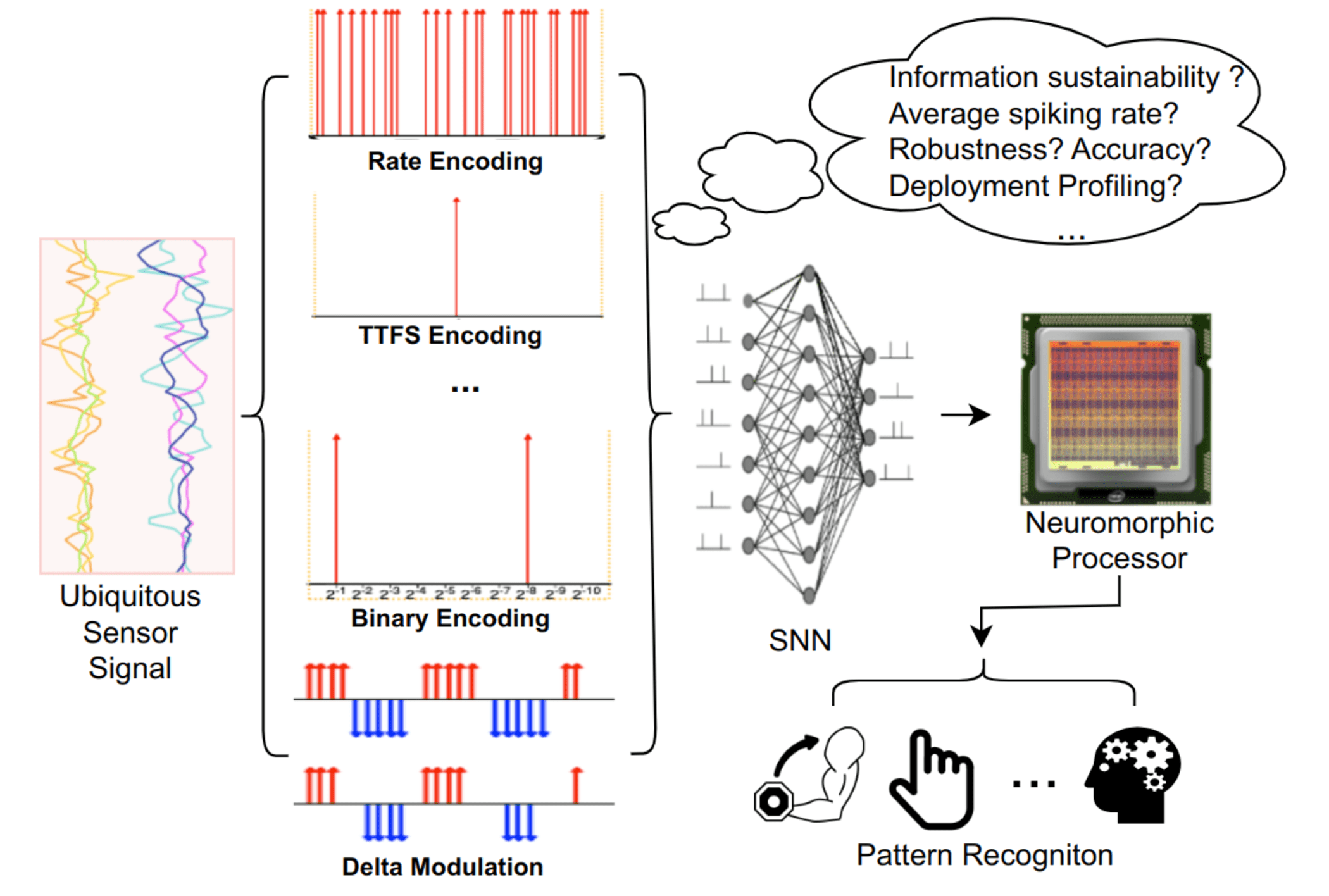}
    \caption{Ubiquitous close-loop neuromorphic embedded system}
    \label{fig:overview}
\end{figure}

Most previous studies on encoding methods have focused on vision data transformation, such as rate encoding (signal represented by the number of spikes within a time window), and delta modulation (spikes generated when signal change exceeds a predefined threshold)~\cite{subbulakshmi2021biomimetic, guo2021neural}. These methods vary in robustness, training process, and deployment performance, and have been utilized in various works~\cite{auge2021survey, kim2022rate}. For example, latency encoding (signal converted to a timestamp within a window) typically offers the best processing latency and energy consumption with fewer synaptic operations while being more susceptible to noise~\cite{oh2022neuron}. Rate coding, found in sensory systems like the visual cortex and motor cortex~\cite{srivastava2017motor}, is high resilience to input noise while limited by a lengthy processing period~\cite{fida2023active, velichko2020concept}. Two-dimensional spike streams from DVS cameras or image-to-spike transformations store information spatially and temporally~\cite{niu2023research, bu2023optimal}, whereas one-dimensional spike stream lack a spatial domain, making them more sensitive and less robust to noise~\cite{el2023efficiency, schoepe2023closed}. The training and deployment performance also differs between one- and two-dimensional spike streams under different encoding schemes. An enlightening preliminary encoding evaluation on ubiquitous sensor signals will pave the way for ubiquitous neuromorphic sensor front-end design, aiming to build real end-to-end neuromorphic computing systems, and further pushing the envelope of energy and latency efficiency of AI embedded edge systems. 
This work explores encoding schemes for time series sensor data in ubiquitous computing (as Fig. \ref{fig:overview} depicts), using inertial measurement unit (IMU) data as a case study. IMUs, commonly found in portable devices, play a dominant role in motion sensing and mimic the vestibular system's inertial sensing for balance and spatial orientation~\cite{day2005vestibular, villani2021modeling}. Previous works have explored biophysical neuromorphic vestibular systems by connecting the IMU to a custom Very Large Scale Integration (VLSI) neuromorphic devices using pulse-frequency modulation to replicate biological dynamics~\cite{corradi2014towards}. IMUs have also been used in robotics for neuromorphic motor control, providing continuous feedback~\cite{blum2017neuromorphic, stagsted2020towards}. Recent work employs rate encoding to derive spike trains from IMU for neuromorphic cognitive perception of motion information~\cite{jiang2023mammalian}. 
Additionally, incorporating neuromorphic technologies into robots from perception to motor control, including event-driven IMUs for closed-loop control, is advocated for fully neuromorphic robotic systems ~\cite{bartolozzi2022embodied}.%In \cite{andreou2013bio}, a micro-fluidic gyroscope that mimics the natural vestibular semicircular canals is presented, which offers significant advantages in terms of power consumption and reliability compared to prior art. Such pioneering studies indicate the value and necessity of further exploring neuromorphic inertial signals, especially in the domain of encoding optimization. 
Since no off-the-shelf neuromorphic-form IMU exists today, this work encodes numerical data from IMU sensors into event-form offline for evaluation. The average firing rate, information loss, signal-to-noise ratio, and the accuracy, robustness, and deployment performance of SNN models under were assessed under different encoding schemes. Specifically, four types of encoding schemes and their variations were evaluated. Time-to-first-spike encoding achieved the lowest firing rate (2\%) and a classification accuracy of 89\%. Rate encoding, with a suitable value-to-probability mapping function, delivered the highest classification accuracy (91.7\%). Binary encoding balanced information reconstruction and noisy spike error resistance. Multi-threshold delta modulation exhibited the best robustness to error spikes (only 0.7\% accuracy drop with 0.1 rate of noisy rate).This work aims to provide the neuromorphic community with an overview of the advantages and limitations of current mainstream encoding schemes for ubiquitous sensor signals.

\section{Related Work}
\label{sec:related_work}

Starting with DVS cameras~\cite{lichtensteiner2008128x128, liu2010neuromorphic}, neuromorphic perception modalities have expanded over the past decades to include audition~\cite{anumula2018feature, liu2010use, liu2013asynchronous}, touch~\cite{rongala2018neuromorphic, oddo2016intraneural, janotte2022neuromorphic}, and olfaction~\cite{vanarse2017investigation, dennler2023limitations}. 
Although vision-based neuromorphic intelligence has impressively developed both academically and commercially~\cite{yao2024spike, rutishauser2023colibries, bian2023colibriuav}, one-dimensional sensing modalities like audio and haptic-based sensing lag behind due to limited application scenarios. However, these studies are rapidly emerging ~\cite{bartolozzi2022embodied, tayarani2021event}, particularly in robotics~\cite{bartolozzi2021neuromorphic, yang2023neuromorphic,zhao2023learning }, and have shown impressive latency and energy performance using SNNs for inferences~\cite{bos2023sub, sharifshazileh2021electronic}, offering valuable insights for more ubiquitous scenarios. 
Table~\ref{relatedwork} highlights several studies exploring SNNs for ubiquitous computing with low-dimensional signals, including their encoding schemes, training approaches, and evaluation metrics. For example, ECG classification using SNNs with different training strategies has been validated on different neuromorphic processors, achieving competitive accuracy~\cite{corradi2019ecg, buettner2021heartbeat, gerber2022neuromorphic}. The authors in ~\cite{corradi2019ecg} employed delta modulation to encode ECG signals into spikes, benefiting from spike sparsity and the on-demand nature of the encoding.

Most researchers use delta modulation to encode one-dimensional sensor signals without comparing different encoding schemes. In contrast, rate encoding is commonly used for two-dimensional event camera data~\cite{wu2019deep, sboev2020solving, han2020deep, mostafa2017fast}. A few comparative studies on encoding schemes were also carried out with experiments on image data~\cite{kim2022rate, auge2021survey, guo2021neural, chen2023hybrid}. 
This work differs by providing a comprehensive performance comparison of encoding schemes on one-dimensional sensory data, which is pervasively generated in ubiquitous embedded systems. Although previous literature shows the importance of encoding schemes on latency and accuracy, there are only two works that evaluate ubiquitous signal encoding~\cite{snr, forno2022spike}. However, \cite{snr} is limited to temporal encoding without evaluating training and deployment results, and ~\cite{forno2022spike} compares the spike encoding techniques but uses a complex method of converting sensor signals into sonograms.
In contrast, our evaluation directly trains SNN from encoded spiking streams under different schemes, avoiding intermediate processes like spike-to-frame and ANN-to-SNN conversion, ensuring the soundness of our conclusions.

\section{Background}
\label{sec:bg}

\subsection{Spike Encoding}\label{theory:encodings}
%This work aims to implement various approaches to encode numerical data into spike trains for information representation. 
Spike encoding takes each single numerical data and produces spike trains consisting of zeros and ones (and negative ones). The length of the spike train for each numerical data is the number of possible spike occurrences \emph{N}. The time between two numerical data points is normally constant and is inferred as the time window. The time difference within the spike train between possible spikes is named the time step ($\delta t$). For some encoding schemes, the spikes are not placed sequentially in the time window but rather create \emph{N} time channels in parallel. Thus, the time step equals the time window. Two pivotal metrics for spike encoding are:
\begin{enumerate}
\item Average Firing Rate (AFR), defined as the number of spikes in a spike train over the total number of possible spike occurrences $N$: $AFR =  \frac{\sum |signal_{spike}|}{N} $, indicating the spike density after encoding. 
%A higher spiking density in the encoded spike train will be more robust to noisy spikes.

%\begin{equation}\label{equ:afr}
%    AFR =  \frac{\sum |signal_{spike}|}{N} 
%\end{equation}

\item Information loss, represented by SNR (Signal to Noise Ratio)\cite{snr} for quantification,
%a general metric for encoding schemes evaluation. In this work, one of the research questions is, if the information loss of an encoding scheme could hint at its performance in training. 
where lower values indicate worse information preservation, is applied: $SNR = 10 * log(\frac{P_{signal}}{P_{err}})[dB]$
%\begin{equation}
%    SNR = 10 * log(\frac{P_{signal}}{P_{err}})[dB] \\
%\end{equation}
with $P_{signal}$ indicates the average power of the original signal and $P_{err}$ menas the average power of the difference between the signal and the reconstruction..
%\begin{itemize}
%    \item $P_{signal}$: The average power of the original signal.
%    \item $P_{err}$: The average power of the difference between the signal and the reconstruction.
%\end{itemize}
\end{enumerate}

At a higher level, two groups of encoding schemes exist, the rate encoding and the temporal encoding~\cite{auge2021survey}. In the following, the main underlying encoding schemes of the two groups are described.

\subsubsection{Rate Encoding}\label{theoy:rate_enc}
also defined as frequency encoding, uses the rate of occurring spikes to encode data.
Converting the normalized signal to firing rates can be done either directly or by using a function mapping the signal to the firing rates, e.g. the CDF (Cumulative Distribution Function) of a random distribution. In direct mapping, the expected number of generated spikes $N_{spikes}$ is decided by the time steps in a spiking window ($\delta t$ $\times$ $N$) and the firing rate, which is equal to the original numerical value in case the signal is already normalized. 
%it is possible to recreate an approximation of the sensor's signal. This means that the number of spikes $N_{spikes}$ for a data point over the total of possible spikes $N$ approximates the numerical value of the original firing rate. 
%In the simple case where the signal is already normalized, the firing rate is equal to the signal's value $v_{signal}$. 
Therefore, reconstructing the signal from the spike train is straightforward when checking the information loss: $v_{signal} \approx \frac{N_{spikes}}{N}$.
%\begin{equation}
%    v_{signal} \approx \frac{N_{spikes}}{N}
%\end{equation}
When using random distributions CDF to convert the values to firing rates, the distribution's PPF (Percent Point Function, inverse of CDF) has to be used to convert the firing rates back to the numerical values.

\subsubsection{Temporal Encoding}\label{theoy:temporal_enc}
Instead of the spike rates, temporal encoding uses the exact timestamp of the spikes to store information. For example, the Time-To-First-Spike (TTFS) encoding encodes the information into a response time. Typically, a higher input value leads to a faster response time. The simplest approach to convert the value into the response time is using the linear relation, and the response time is simply $t_{response} = 1 - s_{value}$, 
%\begin{equation}
%    t_{response} = 1 - s_{value}
%\end{equation}
with $t_{response}$ is the response time and $s_{value}$ is the normalized signal value.
In this work, a logarithmic function is also used for TTFS encoding. The value can be reconstructed by registering the time a spike is fired. 
Another temporal encoding scheme explored in this work is binary encoding, where the time window between two data points is replicated $n_{bits}$ times to form $n_{bits}$ weighted trains.
%divided into a number $n_{bits}$ of frames. 
If a spike fires in a train, the bit is considered a one and vice versa. 
%The signal value is encoded as its binary fraction rounded to the number of bits used by the encoding. 
Reconstructing the signal from binary encoded spike trains is achieved by adding the spikes weighted with their binary fraction value.
The third temporal encoding scheme explored in this work is delta modulation, also named step-forward encoding. It generates a spike when the signal change surpasses a predefined threshold. 
%This spike has either a positive or negative polarity, signifying whether the signal is increasing or decreasing.

\subsection{Spiking Neural Networks}
Different from conventional deep neural networks, SNNs take sparse spike trains as input, and the neurons in the network generate spikes only when the membrane potential exceeds the threshold and reset the potential after spiking, a pattern mimicking the biological neurons. The most widely used idle spiking neuron model is the LIF (Leaky Integrate and Fire) neuron, which has a leaking component for the membrane potential, mimicking the natural dynamic processes of neurons. One variation of it is the CUBA (Current Based LIF) neuron \cite{bouanane2023impact}, which uses current to smooth incoming spikes, leading to a finite rise time in the membrane potential rather than peaking jumps as in the LIF model. 
%As Fig. ~\ref{fig:cuba} depicts, three incoming spikes raise the membrane potential in a finite time; when the threshold is met, the potential drops, and a spike (green dot) is fired; The third spike raises the membrane again, but the threshold is not met, the leaky part can be seen decreasing the membrane potential over time. 
%In this work, the CUBA neuron model is used in the spiking neural network. 
%It has been shown to outperform the LIF~\cite{cuba}.

%\begin{figure}
%    \centering
%    \includegraphics[width=1.0\linewidth]{Figures/cuba.png}
%    \caption{CUBA neuron: the finite rising periods are the neuron's response to incoming spikes. The two dots signify that the neuron fired a spike at the time.}
%    \label{fig:cuba}
%\end{figure}

\subsection{Dataset}\label{dataset}

We choose a challenging IMU-based gym activity recognition dataset (RecGym\cite{RecGym}) in this work to explore the performance of each encoding as a case study. The magnitude is unevenly distributed in the normalized dataset with a majority magnitude near the median; thus, compared to other IMU-based datasets, the encoding needs to be more carefully tuned to represent the raw information and decrease the information loss.  
%The dataset is a publicly available labeled data set, recording gym activities with sensing units composed of the IMU and the human body capacitive (HBC) sensor \cite{bian2019passive}, with a frequency of 20$Hz$. 
%The sensing units were worn at three positions: on the wrist, in the pocket, and on the calf, sampling the motion context with a frequency of 20$Hz$. 
%The to be recognized twelve activities were recorded, including eleven workouts: Adductor, ArmCurl, BenchPress, LegCurl, LegPress, Riding, RopeSkipping, Running, Squat, StairsClimber, Walking, and a "Null" activity when the volunteer hung around between different workouts session. 
%Each participant performed the above-listed workouts for five sessions in five days, and each session lasted around one hour. Altogether, fifty sessions of gym workout data from ten volunteers were presented. 
A detailed description of the dataset can be found in \cite{bian2022contribution}, in which the authors deployed a Resnet neural network for workout recognition and achieved an accuracy of 91\%. 

\section{Encoding Implementation}

%As explained in section~\ref{theory:encodings}, there are two main categories on a higher level to encode numerical data as spikes: rate encoding and temporal encoding. In this work, different specific encoding schemes are implemented and compared. For the rate encoding, different probability distributions for signal-to-rate mapping are explored. For the temporal encoding, a multi-threshold delta modulation approach, a phase-dependent binary representation, and a latency-based approach are explored. 
%Because the used framework Lava~\cite{lava} supports negative flanking spikes, they were introduced to increase the encoding's functionality.

\subsection{Rate Encoding}\label{rate_enc}
With rate encoding, a numerical data point is represented by the number of spikes within a time window. For sensor data, rate encoding generates a spike train with the same duration as the signal's sampling interval. We first use different CDF to map the value to the firing rate aiming to decrease the information loss, and then fire the spike within the time window with a Bernoulli trial, where the number of trials is $n$ = 1 for each time step, and the probability of a successful spiking($P_{fr}$) is the mapped firing rate with CDF mapping ($F$): $v_{signal} \sim B(n,P_{fr} ), P_{fr} = F(v_{signal}) $.
%\begin{equation}
%    v_{signal} \sim B(n,P_{fr} ), P_{fr} = F(v_{signal}) 
%\end{equation}
%\begin{equation}
%    P_{fr} = F(v_{signal}) 
%\end{equation}
As rate encoding uses random decisions, the returned spike trains are not deterministic. The ideal firing spike number should be close to the product of the mapped firing rate and number of spike possibilities.
In the following subsections, the probability mapping functions and their implementations are described.

\subsubsection{Uniform Distribution}

\begin{figure}
    \centering
    \includegraphics[width=1.0\linewidth]{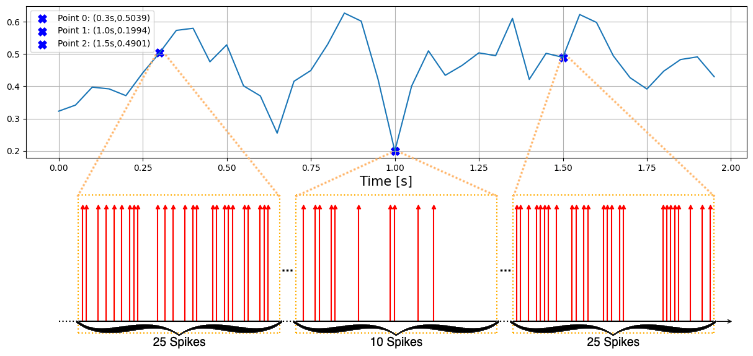}
    \caption{Rate encoding for 3 data points using a uniform distribution. A higher value yields more spikes.}
    \label{fig:bin_enc}
\end{figure}

\begin{algorithm}[]
  \SetAlgoLined
  \KwInput{$signal$}
  \KwResult{$spike\_train$.}
  initialize an empty $spike\_train$\;
  \For{m \textbf{in} length of $signal$}{
     $v_{signal}$ = $signal[m]$\; 
     $P_{fr} = uniform.cdf(v_{signal})$ \;
     \For{n \textbf{in} length of time steps for each $v_{signal}$ (50)}{
        $spike\_train[50m + n] = B(1,P_{fr})$
     }
  }
  return $spike\_train$ \;
  \caption{Rate Encoding using uniform distribution function}
  \label{algo:uniform}
\end{algorithm}

The most straightforward rate encoding approach is to consider the real value as the firing rate of the spike train directly. With the value as the firing rate, a Bernoulli trial is passed through all possible spiking time steps with the value as the success possibility. The time step $\delta t$ between possible spikes is set to one millisecond. This means each individual data point in a 20Hz signal has 50 possible spike occurrences. Figure~\ref{fig:bin_enc} shows how three different data points are encoded into spikes. 
Though uniform distribution-based rate encoding is easy to implement, one problem is that the resolution of the encoding is relatively low at $0.02$, especially considering that most of the signals deviate not too much around $0.5$. This means that there might be practically no difference between the encoded signals of different workouts and the classification becomes worse. 
%One simple solution would be to increase the number of spikes. 
%Due to memory concerns and because more spikes lead to a higher memory consumption, different approaches were considered.

\begin{figure}
    \centering
    \includegraphics[width=1.0\linewidth]{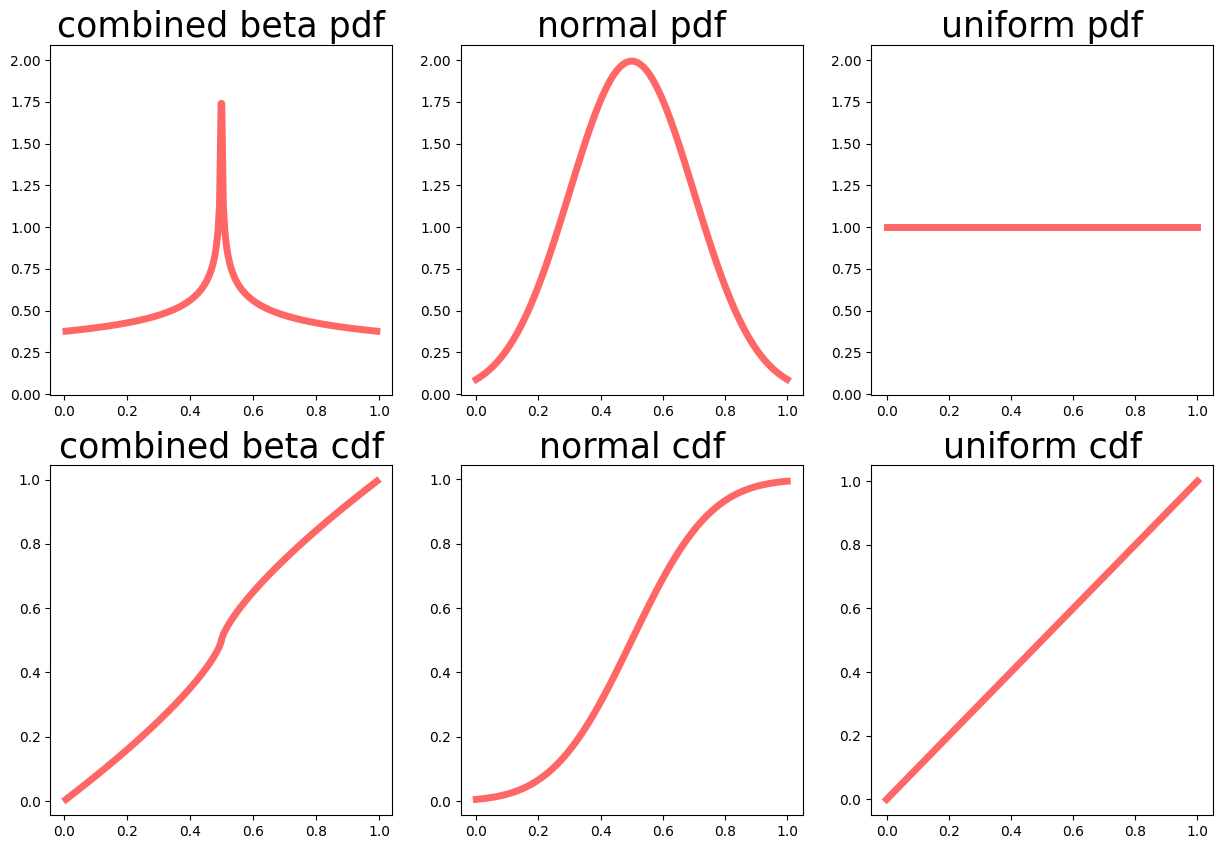}
    \caption{Comparing the three distributions: the combined beta distribution better captures varying values around the middle point around 0.5 (PDF: Probability Density Function, CDF: Cumulative
Distribution Function}
    \label{fig:beta_norm_uniform}
\end{figure}

\begin{algorithm}[]
  \SetAlgoLined
  \KwInput{$signal$}
  \KwResult{$spike\_train$.}
  initialize an empty $spike\_train$\;
  \For{m \textbf{in} length of $signal$}{
     $v_{signal}$ = $signal[m]$\; 
     $P_{fr} = normal.cdf(v_{signal})$ with $\mu=0.5$ and $\sigma^2=0.2$\;
     \For{n \textbf{in} length of time steps for each $v_{signal}$ (50)}{
        $spike\_train[50m + n] = B(1,P_{fr})$
     }
  }
  return $spike\_train$ \;
  \caption{Rate Encoding using the normal distribution function}
  \label{algo:norm}
\end{algorithm} 
\subsubsection{Normal Distribution Function}
\begin{figure}
    \centering
    \includegraphics[width=1.0\linewidth]{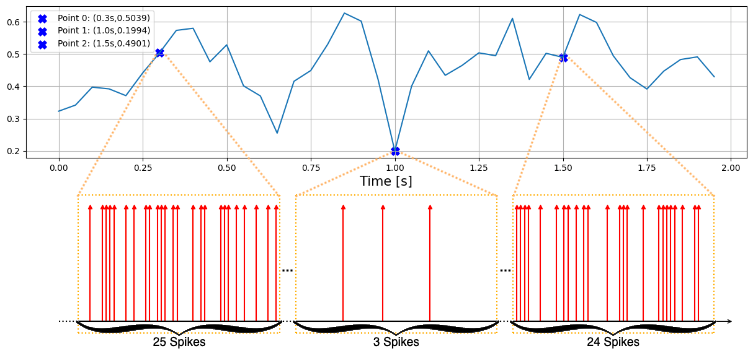}
    \caption{Rate encoding for 3 data points using the normal distribution function.}
    \label{fig:norm_enc}
\end{figure}

To achieve a more sensible resolution around the midpoint, two different probability distribution functions are chosen, as Figure~\ref{fig:beta_norm_uniform} shows. The Probability Density Function (PDF) in the top row is the derivation of the Cumulative Distribution Function (CDF). As this derivation increases, the more perceptive the function's output reacts to the input. The algorithm for the rate encoding scheme using the normal distribution~(Algorithm \ref{algo:norm}) is conceptually the same as the encoding scheme using the uniform distribution~(Algorithm \ref{algo:uniform}). The only difference is that the value-to-rate mapping is changed using the CDF of the normal distribution. 
%The standard deviation is set to 0.2, as the resulting distribution is expected to generate no spikes at zero. On top of that, its PDF curve fits well~\ref{fig:beta_norm_uniform}.
As Figure~\ref{fig:norm_enc} shows, in contrast to using the uniform distribution function, spike counts are lower for values below 0.5 and higher for higher values, respectively, supplying more sensitivity near 0.5.
However, the drawback of using a normal distribution function is that the probabilities change slowly near the endpoints zero and one, potentially resulting in inaccurate representation. 

%\begin{figure}
%    \centering
%    \includegraphics[width=1.0\linewidth]{Figures/beta_dist.png}
%    \caption{Beta distribution showing that values close to one result in more distinguishable probabilities, while values at zero still have decently differing probabilities.}
%    \label{fig:beta_dist}
%\end{figure}

\subsubsection{Combined Beta Distribution}
\begin{figure}
    \centering
    \includegraphics[width=1.0\linewidth]{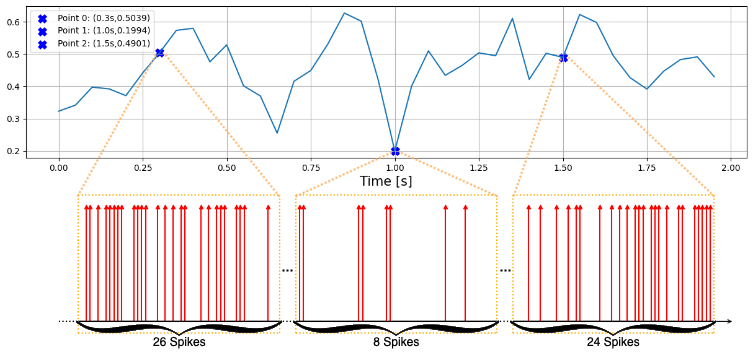}
    \caption{Encoding scheme using the combined beta distribution. The intervals for different numbers of spikes are very tight, around 0.5.}
    \label{fig:beta_enc}
\end{figure}

To further exploit the signal feature, a third distribution is introduced. The idea is to apply a distribution whose CDF has a high slope around the median value and still gives sensitive probabilities at the other side. The beta distribution with its parameters $\alpha$ and $\beta$ can do this very well for the endpoints zero and one. 
%Figure~\ref{fig:beta_dist} shows that for the parameters $\alpha=1$ and $\beta<1$, the CDF gives discernible differences in probabilities for only small differences in values close to zero. 
%\begin{align*}
%    pdf_{beta}(x) &= \frac{1}{B(\alpha,\beta)} * x^{\alpha-1} * (1-x)^{\beta-1} \\
%    B(\alpha,\beta) &= \int^0_1 u^{\alpha-1} * (1-u)^{\beta-1} du
%\end{align*}
Setting $\alpha$ to one creates the peak at one, and $\beta<1$ gives the curvature of the PDF. To apply this distribution to the dataset in this work, the beta encoding is applied twice. Once for values between 0 and 0.5 with $\alpha=1$ and $\beta=0.75$ and once for values between 0.5 and 1 with the parameters reversed, so $\alpha=0.75$ and $\beta=1$. This gives the distribution in Figure~\ref{fig:beta_norm_uniform}. The algorithm is slightly different from the previous ones, and the data point is converted to the firing rate by using the CDF of the combined beta distribution function (Algorithm \ref{algo:beta}, \ref{algo:combined_beta.cdf}). 
It can be seen in Fig.~\ref{fig:beta_enc} that the rate encoding utilizing the combined beta distribution can distinguish small differences in values close to 0.5 even better than with the normal distribution. Meanwhile, it generates a higher count of spikes at the signal's low point.

%So converting the firing rates and using samples from the uniform distribution provided by NumPy~\cite{numpy} was considerably faster.

\begin{algorithm}[]
  \SetAlgoLined
  \KwInput{$signal$}
  \KwResult{$spike\_train$.}
  initialize an empty $spike\_train$\;
  \For{m \textbf{in} length of $signal$}{
     $v_{signal}$ = $signal[m]$\; 
     $P_{fr} = combined\_beta.cdf(v_{signal})$\;
     \For{n \textbf{in} length of time steps for each $v_{signal}$ (50)}{
        $spike\_train[50m + n] = B(1,P_{fr})$
     }
  }
  return $spike\_train$ \;
  \caption{Rate Encoding using the combined beta distribution}
  \label{algo:beta}
\end{algorithm}

\begin{algorithm}[]
  \SetAlgoLined
  \KwInput{$v{signal}$}
  \KwResult{Return the $P_{fr}$.}
  \eIf{$v{signal}<0.5$}
{
    $v_{normalized} = 2*v{signal}$ \;
    $P_{fr} = beta.cdf(v{normalized}, \alpha=1, \beta=0.75)$ \;
}{
    $v_{normalized} = 2*v{signal}-1$ \;
    $P_{fr} = beta.cdf(v{normalized}, \alpha=0.75, \beta=1)$ \;
}
  return $P_{fr}$ \;
  \caption{Values are converted to firing rates by the combined beta distribution($combined\_beta.cdf()$)}
  \label{algo:combined_beta.cdf}
\end{algorithm}

%The function $combined\_beta.cdf$ (Algorithm \ref{algo:combined_beta.cdf}) is the combination of the two beta distributions with the pair $\alpha=1$, $\beta=0.75$ and $\alpha=0.75$, $\beta=1$, respectively. If the value is smaller than 0.5, the value is broadcast to the interval from 0 to one, and the first distribution is used to get the probability, which is halved to get the firing rate. If the value is greater than 0.5, it is broadcast to the same interval, but now the second distribution is used. The obtained probability is halved again and added to 0.5, to get a firing rate between 0.5 and 1. 

\subsection{Temporal Encoding}

%Temporal encoding indicates that the presence of a spike at a certain time step holds the information. In contrast to the rate encoding, the different temporal encoding schemes need specific functions for implementation.

\subsubsection{Time-To-First-Spike (TTFS)}
\begin{figure}
    \centering
    \includegraphics[width=1.0\linewidth]{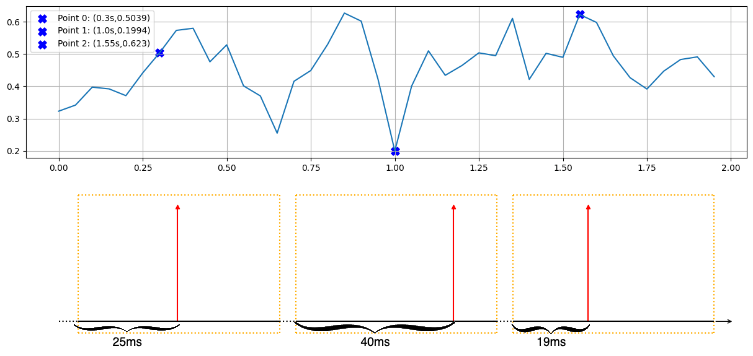}
    \caption{TTFS encoding, showing how greater values fire a spike earlier than smaller ones.}
    \label{fig:ttfs_linear}
\end{figure}

TTFS is one of the classic time-dependent encoding schemes. The value of the original signal is converted to a timestamp within a window. The time passed within the window until the spike appears encodes the numerical value. The greater the value, the faster the spike appears.
As Fig. ~\ref{fig:ttfs_linear} shows, the firing time is linearly dependent on the numerical value, e.g. a value of 0.5 will fire a spike after 25ms. 

In this work, the function implemented to encode a signal using TTFS takes the signal as input, as well as a function parameter, which indicates whether a linear or logarithmic function should be used to calculate the response time of the spike. 
The TTFS-encoded spike train has the same dimensions but with the length in the time dimension multiplied by the number of time steps.
%The linear TTFS encoding is implemented by Algorithm \ref{algo:ttfs_linear}.
%first calculating the spike time of all data points and then multiplying it by the number of time steps (Algorithm \ref{algo:ttfs_linear}). 

%\begin{algorithm}[]
%  \SetAlgoLined
%  \KwInput{$signal$}
%  \KwResult{$spike\_train$.}
%  initialize an empty $spike\_train$\;
%  \For{m \textbf{in} length of $signal$}{
%     $v_{signal}$ = $signal[m]$\; 
%     $n = int(v_{signal} \times TimeSteps(50))$\; 
%      $spike\_train[50m + n] = 1$\; 
%  }
%  return $spike\_train$ \;
%  \caption{Linear TTFS encoding}
%  \label{algo:ttfs_linear}
%\end{algorithm}

%\begin{figure}
%    \centering
%    \includegraphics[width=1\linewidth]{Figures/ttfs_neg.png}
%    \caption{Comparison between TTFS encodings using negative spikes. Spikes tend to occur sooner for negative spikes.}
%    \label{fig:ttfs_neg}
%\end{figure}

\begin{figure}
    \centering
    \includegraphics[width=1.0\linewidth]{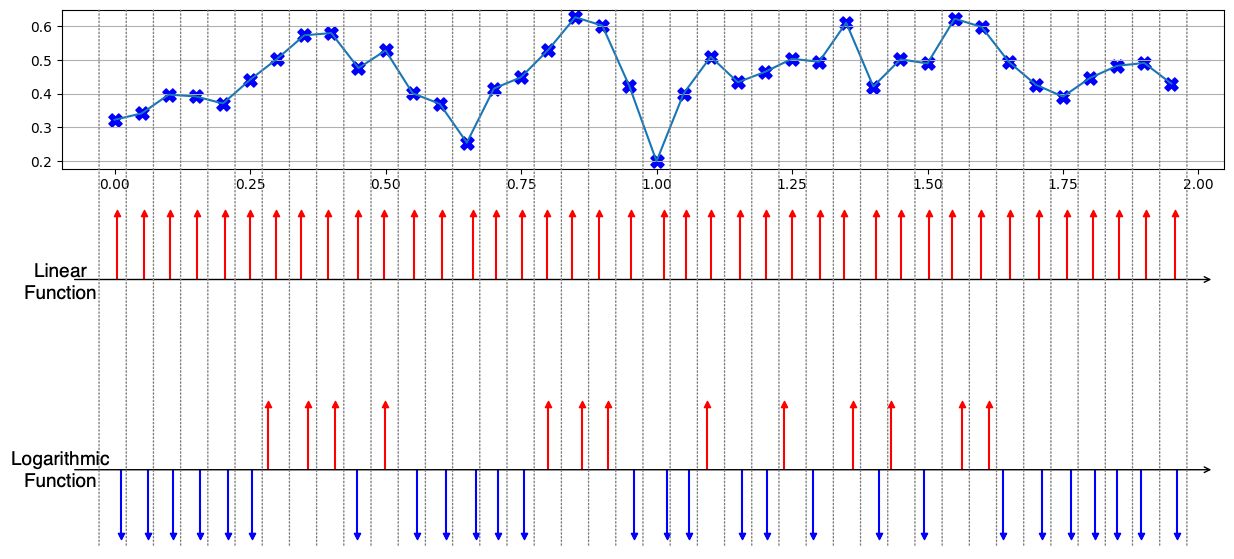}
    \caption{The full spike trains for the TTFS encoding are shown. 
    %It has to be mentioned, that TTFS encodings only use one spike per time window, which makes it attractive for energy concerns.
    }
    \label{fig:ttfs_lin_log}
\end{figure}

To address the data set characteristic of having many signals staying close to 0.5, a logarithmic function %(Fig. ~\ref{fig:ttfs_log_func}) 
is considered besides the linear one. However, this still didn't solve the problem completely, as the sensitivity is only higher near zero. In this case, instead of transforming the raw data, we take the difference to 0.5 as the encoding object. For data less than 0.5, a negative spike train is then used for a better representation. Algorithm \ref{algo:ttfs_log_neg} explains the detailed encoding process, and Fig. ~\ref{fig:ttfs_lin_log} shows an example of a full spike train for the whole signal.

%\begin{figure}
%    \centering
%    \includegraphics[width=1.0\linewidth]{Figures/ttfs_log_func.png}
%    \caption{The logarithmic function used by the TTFS encoding $y=-20\log_{10}(x)$.}
%    \label{fig:ttfs_log_func}
%\end{figure}

\begin{algorithm}[]
  \SetAlgoLined
  \KwInput{$signal$}
  \KwResult{$spike\_train$.}
  initialize an empty $spike\_train$\;
  \For{m \textbf{in} length of $signal$}{
     $v_{signal}$ = $signal[m]$\; 
     $diff = 2*(v_{signal}-0.5) $ \;
     \eIf{$diff<0$}{
            $n =  int( -20\log_{10}(-diff))$\;
            $spike\_train[50m + n] = -1$\; 
        }{
            $n=  int(-20\log_{10}(diff))$\;
            $spike\_train[50m + n] = 1$\; 
        }
  }
  return $spike\_train$ \;
  \caption{The logarithmic function used by the TTFS encoding.}
  \label{algo:ttfs_log_neg}
\end{algorithm}

%\begin{algorithm}[]
%  \SetAlgoLined
%  \KwInput{$signal$, $t_{steps}$, $threshold$}
%  \KwResult{Return the $spike\_train$.}
%  initialize an empty $spike\_train$ \;
%  $\tau = (t_{steps}-1)*log(threshold)$ \;
%    \For{all data point values $v_{signal}$}{
%        $diff = 2*(v_{signal}-0.5) $ \;
%        $spike_{time} =  round\_up(\tau * log(diff))$\;
%        \eIf{$diff<0$}{
%            $spike\_train[data\_point + t_{steps} - spike_{time}] = \textbf{-1}$ \;
%        }{
%            $spike\_train[data\_point + t_{steps} - spike_{time}] = \textbf{1}$ \;
%        }
%    }
%  return $spike\_train$ \;
%  \caption{The logarithmic function used by the TTFS encoding.}
%  \label{algo:ttfs_log_neg}
%\end{algorithm}

%\begin{figure}
%    \centering
%    \includegraphics[width=1.0\linewidth]{Figures/ttfs_lin_log_neg.png}
%    \caption{Comparison between TTFS encodings with different functions. The logarithmic function is more sensitive around 0.5, but the spikes fire later.}
%    \label{fig:ttfs_neg_lin_log}
%\end{figure}

\subsubsection{Binary Encoding}

\begin{figure}
    \centering
    \includegraphics[width=1.0\linewidth]{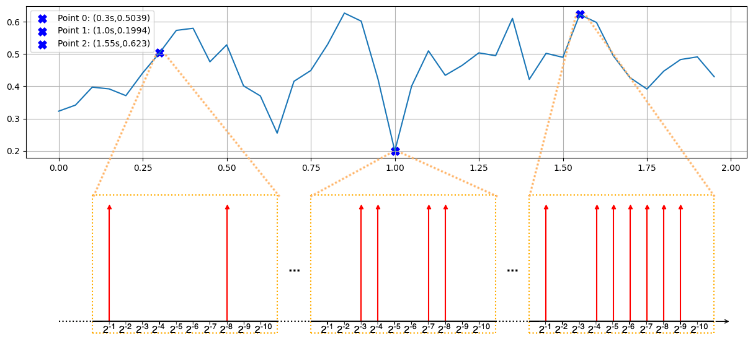}
    \caption{Binary encoding using ten bits to encode the numerical value. 
    %To be noticed, here we plot the spikes in a sequential way within one train to save space. In the implementation, ten trains are generated in parallel.
    }
    \label{fig:bin_10}
\end{figure}

Binary encoding works by mapping the numerical value to a number of bits. In this scheme, both the number of spikes and the timestamp at which the spikes occur are relevant for information encoding. To get the spike train, a data point is represented by its binary fraction. The number of bits decides how many fractions are used. Algorithm \ref{algo:binary} summarizes the implementation. %The value of the data point is compared to the first fraction of $\frac{1}{2}$. If the fraction is smaller than the remaining number, it is subtracted from the number, and a spike is generated. If the fraction is bigger, no spike is generated. Then the fraction is divided by two and so on for a number of rounds equal to the number of bits provided (Algorithm \ref{algo:binary}). 
By doing the encoding this way, it is ensured that an input number of one generates spike trains containing all ones where the train number is equal to the number of bits. 
%This means to convert the spike train back to an approximation of the original value, the fraction, where a spike was generated only need to be added together, resulting in the approximation. 
Figure~\ref{fig:bin_10} shows a ten-bit binary encoding, giving a resolution of $\frac{1}{2^{10}} \approx 0.001$. This is more than the resolutions of the previous linear TTFS and uniformly distributed rate encodings. Thus, the data set is also encoded using six bits, to compare the binary encoding with the other encodings sharing similar resolutions.

\begin{algorithm}[]
  \SetAlgoLined
  \KwInput{$signal$, $n\_bits$}
  \KwResult{Return the $spike\_train$.}
  initialize an empty $spike\_train$ \;
    \For{m \textbf{in} length of $signal$}{
        $v_{signal}$ = $signal[m]$\; 
        $bit\_decision = 0.5$ \;
        \For{$bit$ \textbf{in} $range(n\_bits$)}{
            \If{$v_{signal} - bit\_decision>0$}{
                %$spike\_train[m * n\_bits + bit] = 1$ \;
                $spike\_train[m, bit] = 1$ \;
                $v_{signal} - = bit\_decision$ \;
            }
            $bit\_decision *= 0.5$ \;
        }
    }
  return $spike\_train$ \;
  \caption{The binary encoding scheme.}
  \label{algo:binary}
\end{algorithm}

%As Algorithm~\ref{algo:binary} explains, the spikes are not stored as a one-dimensional time-dependent spike train, but in a second dimension. This way the binary encoded data or more precisely the bits themselves can be fed to their own respective input nodes in the SNN. Fig.~\ref{fig:bin_6} shows how this works for the number of bits set to six.

%\begin{figure}
%    \centering
%    \includegraphics[width=1.0\linewidth]{Figures/binary_6.png}
%    \caption{Fully encoded signal using 6 bit binary encoding. The bit spike trains are stored in another dimension and not in the time dimension.}
%    \label{fig:bin_6}
%\end{figure}

%This makes the encoding scheme arguably no longer fit into the temporal category. But because a neuromorphic spike sensor generating the spikes using binary encoding would probably still send them through one channel as a time-dependent spike train, this can be considered a data conversion to use the SNN. Therefore, we argue that this version of the binary encoding still fits into the temporal encoding family.

\subsubsection{Multi-Threshold Encoding}
\begin{figure}
    \centering
    \includegraphics[width=1.0\linewidth]{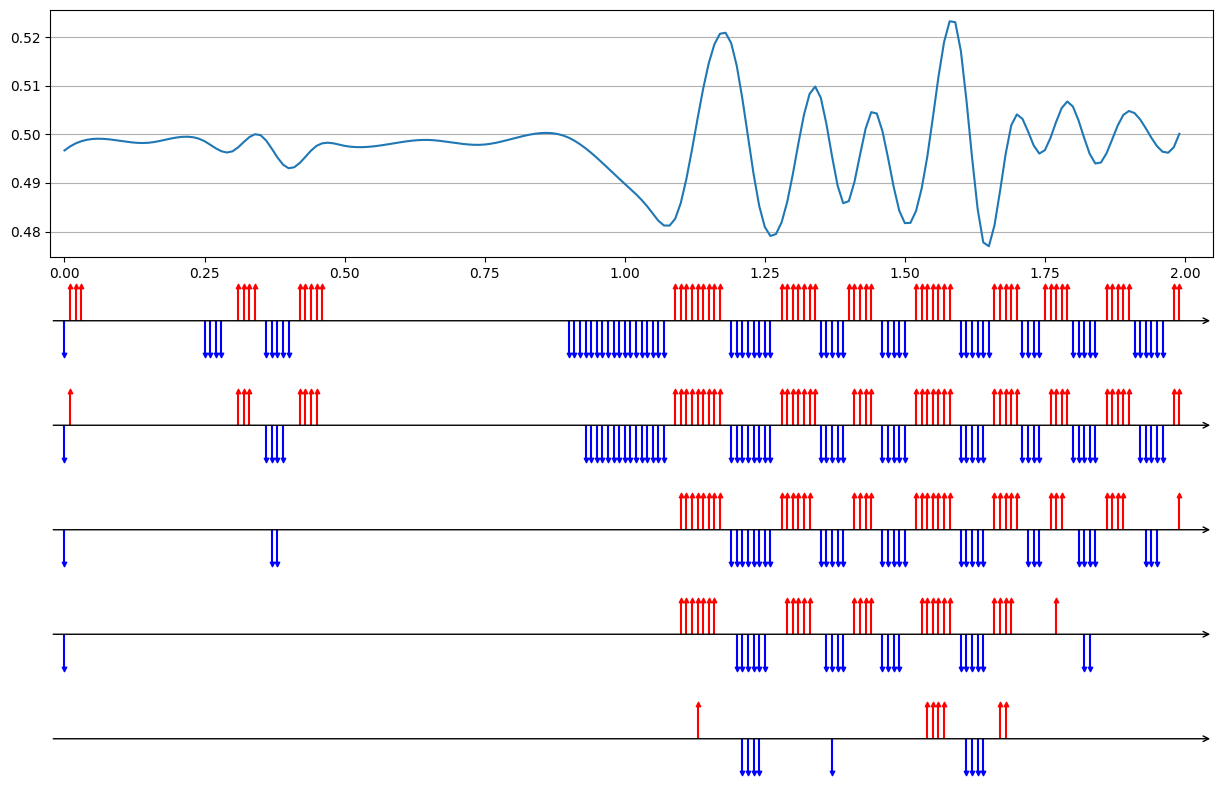}
    \caption{The 5-Threshold encoding scheme. 
    %If a threshold is exceeded between 2 data points, a spike is generated. 
    The five rows are five increasing thresholds.}
    \label{fig:5_threshold}
\end{figure}

%The last explored encoding scheme is multi-threshold delta modulation encoding. T
he main idea is to take the difference between two subsequent data points and check whether a threshold is exceeded to generate positive or negative spikes.
%, depending on if the difference is positive or negative. 
This study empirically applies five thresholds in parallel, as Algorithm \ref{algo:5_threshold} explains. The five thresholds have their own channel as seen in Fig.~\ref{fig:5_threshold}. As the original signals are sampled with 20Hz, to maintain a comparative classification accuracy, the original signals are interpolated to 100 $Hz$. Defining the thresholds is challenging. 
%A small threshold always fires, and a large threshold never fires. On top of that, 
If a signal rapidly increases with a high slope, all thresholds will fire without adequately capturing the change. And if a signal decreases or increases slightly such that the smallest threshold never fires, it becomes indistinguishable from a constant signal. The choice of the threshold values was done experimentally through a set of trials, with recognition accuracy as the criterion.

\begin{algorithm}[]
  \SetAlgoLined
  \KwInput{$signal$}
  \KwResult{Return the $spike\_train$.}
  initialize an empty $spike\_train$ \;
    \For{m \textbf{in} (length of $signal - 1$)}{
        %$v_{signal}$ = $signal[m]$\; 
        $diff = signal[m+1] - signal[m]$ \;
        \For{$i$ \textbf{in} Thresholds $T_i$}{
            \If{$diff > T_i$}{
                $spike\_train[m, i] = 1$ \;
            }
        }        
    }
  return $spike\_train$ \;
  \caption{The 5-Threshold encoding scheme using data modulation.}
  \label{algo:5_threshold}
\end{algorithm}

\section{SNN Implementation}
\label{sec:snn}

With the conversion of numerical sensor data to spike trains via the various spike encoding schemes implemented in last section, an SNN is directly trained for the workout classification. This section details the SNN implementation.

\subsection{Network}\label{imp:network}

In this work, a simple network composed of three hidden dense layers (256-64-12) is trained for the classification. %The first layer's input size needed to be adapted for the 2-dimensional spike trains and whether negative spikes are used. 
The network's output layer had twelve nodes, representing the twelve possible workout labels.
The neurons used for the network are CUBA neurons with three main parameters that need to be set: neuron threshold, the fraction of current decay per time step, and the fraction of voltage decay per time step.
%\begin{itemize}
%    \item Threshold: neuron threshold
%    \item Current\_decay: the fraction of current decay per time step
%    \item Voltage\_decay: the fraction of voltage decay per time step
%\end{itemize}
Further hyperparameters of the network are the dropout probability, the learning rate and the batch size as well as the true and false rate for the loss function. %All these parameters add up and make hyperparameter optimization very challenging.
%Adding to that is, that the optimal value for some of these parameters seem to vary between different spike encoding scheme.
A random Search is performed supported by the WANDB platform to find the optimal network hyperparameters for each encoding scheme.
%In the end, good parameters were found for the encodings using negative spikes. Unfortunately, this was not the case for the encodings using only positive spikes, see the chapter~\ref{discussion} for details.
The framework used in this work is the Lava Software Framework~\cite{lava}, an open-source software framework facilitating neuromorphic computing, ranging from creating and training SNN to deploying them on neuromorphic hardware, the Loihi platform. 
%It has a Python interface, and in the back-end, the performance is accelerated with underlying C/C++/CUDA code.
%The main library used in this work is the Lava-DL library, which contains a variety of tools that enable training and inference for various event-based networks. One of them is SLAYER 2.0, the progression of SLAYER~\cite{Slayer}, a training mechanism for learning synaptic weights and axonal delays overcoming the problem of non-differentiability of the spike function, based on a temporal credit assignment policy for backpropagating error to the preceding layer. 

\subsection{Parameters Setup}
\label{ssec:setup}

\begin{table}[h]
    \centering
    \caption{The chosen thresholds and the AFR they produce in a five-threshold delta modulation encoding scheme.}
    \label{tab:threshold}
    \begin{tabular}{|p{0.6cm}|p{1.2cm}|p{0.8cm}|p{0.8cm}|p{0.8cm}|p{0.8cm}|p{0.8cm}|}
     \hline
     \multirow{2}{1cm}{IMU} & Threshold & 0.0004 & 0.0008 &0.0016 &0.0032 & 0.0064 \\ 
     & AFR [\%] & 66.989 & 53.739 & 38.314 & 23.204 & 11.551 \\
     \hline
     \hline
     \multirow{2}{1cm}{HBC} & Threshold & 0.0001 & 0.0002 & 0.0004 & 0.0008 & 0.0016 \\ 
     & AFR [\%] & 56.835 & 46.496 & 36.121 & 26.818 & 19.383  \\
    \hline
    \end{tabular}
    
\end{table}

\begin{table*}
    \centering
    \begin{threeparttable}
    \caption{Evaluation of different encoding schemes.}
    \begin{tabular}{|p{0.6cm}|p{1.2cm}|p{1.1cm}|p{0.6cm}|p{0.8cm}|p{0.6cm}|p{0.9cm}|p{0.5cm}|p{0.5cm}|p{0.5cm}|p{1.4cm}|p{1.3cm}|}
    %\begin{tabular}{|c|c|c|c|c|c|c|c|c|c|c|}
    \hline
    \multicolumn{2}{|l|}{\centering Encoding} &  \centering Tensor &  \centering Time  & \centering AFR  & \centering SNR  & Accuracy &  \multicolumn{3}{l|}{\centering Robustness \tnote{a}} & \multicolumn{2}{l|}{\centering Deployment Performance \tnote{b}} \\
    \cline{8-12}
     \multicolumn{2}{|l|}{\centering Scheme} &  \centering  Shape (2s) &  \centering  Step (ms) & \centering (\%)  &  \centering (dB) &  & 0.001 & 0.01 & 0.1 & Dynamic Energy(mJ) & Execution Time(ms) \\
    \hline
    %\centering Uniform & &  & &  &  &  &  &  &  & \\
    \multirow{3}{*}{Rate} & Uniform & (1,7,2000) & 1 & 49.943 & 91 & 85.4   &  0.4 & 1.3  & 11.1  & 436.51  & 73.96\\
    & Normal & (1,7,2000) & 1 & 49.912 & 99 & 90.9  & 0.5 & 3.4 & 10.6 & 402.14 & 78.07\\
    & Beta  & (1,7,2000) & 1 & 49.926 & 101 &  \textbf{91.7} & 0.4 & 1.9 & 9.5  & 250.15 &  75.75\\
    \hline
    \multirow{2}{*}{TTFS} & Linear  &  (1,7,2000) & 1 &  \textbf{2} & 74 & 89.1 & 4.1 & 20.7 & 37.3  & 144.39 & 78.29 \\
    & Logarithmic & (1,7,2000)& 1 & \textbf{2} & 87 & 89.2 & 2.5 & 14 & 22.9 & 590.23 &  80.49 \\
    \hline
    \multirow{2}{*}{Binary} & 6 Bits & (6,7,40)& 50 & 33.327 & 88  &  86.5 & 0.7 & 0.9 & 2.5 & 8.87 & 88.75\\
    & 10 Bits & (10,7,40)& 50 & 46.910 & \textbf{139}  & 89.6 & 0.2 & 0.6 & 1.0 & 6.31 & 84.70 \\
    \hline
    \multicolumn{2}{|l|}{Delta Modulation } & (5,7,200) & 10 & 38.527 & 84 & 89.8 & \textbf{0.1} & \textbf{0.2} & \textbf{0.7} & 24.47 & 82.94 \\
    \hline
    \end{tabular}
    \begin{tablenotes}
    \setlength{\columnsep}{0.8cm}
    \setlength{\multicolsep}{0cm}
    %\begin{multicols}{1}
            \item[a] Indicated by the accuracy drops when giving different error probabilities for each input event.
            \item[b] All measurements were obtained using Nx SDK version — on Nahuku 32 board ncl-ghrd-04. The same set of test samples is given for each encoding scheme, and the averaged performance is reported here. 
    %\end{multicols}
    \end{tablenotes}
    
    \label{tab:result_all}
    \end{threeparttable}

\end{table*}

The raw data is forwarded to encoding pipelines together with the following parameters: the time step length is set to one millisecond for all encodings that generate a single spike train per channel, namely the TTFS and all the rate encodings; The variance of the normal distribution is set to 0.2. This is due to a few real tests showing that the information loss of the normal distributed rate encoding is minimal, with a variance of around 0.2; We set the first parameter of beta distribution to one to maintain the CDF curve shape, and meanwhile, considering both the information loss and sensitivity of all data points for rate generation, an empirical value of 0.75 is set; For the binary encoding, the bit numbers six and ten are chosen. Six bits is chosen as $\frac{1}{64}$ gives a nearest resolution to the rate encodings $\frac{1}{t\_steps}$ where $t\_steps = 50$. And ten bits is also explored since the binary encoding's strength of a high resolution for comparably small array sizes is worth exploring;  Table~\ref{tab:threshold} shows the chosen thresholds and their AFR(Average Firing Rate) for multi-threshold delta modulation encoding, indicating that the vast majority of differences are captured. 

%\begin{itemize}
%    \item The time step length is set to one millisecond for all encodings that generate a single spike train per channel, namely the TTFS and all the rate encodings. 
%    \item The variance of the normal distribution is set to 0.2. This is due to a few real tests showing that the information loss of the normal distributed rate encoding is minimal, with a variance of around 0.2.
%    \item We set the first parameter of beta distribution to one to maintain the CDF curve shape, and meanwhile, considering both the information loss and sensitivity of all data points for rate generation, an empirical value of 0.75 is set.
%    \item For the binary encoding, the bit numbers six and ten are chosen. Six bits is chosen as $\frac{1}{64}$ gives a nearest resolution to the rate encodings $\frac{1}{t\_steps}$ where $t\_steps = 50$. And ten bits is also explored since the binary encoding's strength of a high resolution for comparably small array sizes is worth exploring.

%All the above-mentioned encoding schemes are run with both enabled and disabled negative spikes, except for the multi-threshold encoding, as it is the only one, that only works if negative spikes are used. 
%   \item Table~\ref{tab:threshold} shows the chosen thresholds and their AFR(Average Firing Rate) for multi-threshold delta modulation encoding, indicating that the vast majority of differences are captured. 
%\end{itemize}

\subsection{Training}\label{imp:training}
A dataset python class is implemented to load the spike trains from the binary files that contain the encoded spikes and the corresponding workout label. On initialization, the dataset is provided with the sampling time used in the encoding scheme and the signal's dimension. As the spikes are read out as SLAYER events, the shape is composed of: the number of sensor channels (seven in this work), the number of spike trains (e.g. five in a five-threshold delta modulation), 0 and 1/-1 depending on whether spike exists, and the length of the signal (2s in this work). Using SLAYER, the training was done one on a \emph{NVIDIA GeForce RTX 3080 Ti} GPU on the server. The optimizer used is \emph{ADAM}, and the loss function is the SLAYER native loss function SpikeRate, meaning that the event rate in the output layer over the whole duration was taken as the confidence score. The network, loss function, optimizer, and classifier were given to the SLAYER assistant, which bundles the training and testing workflows. 
The training epoch was set to 100, the achieved train and test accuracy was stored during training, and the best-performing weights were then exported.

\section{Results}\label{results}

Besides the previously described AFR and SNR, the following metrics are also used to evaluate each encoding scheme:
\begin{itemize}
    %\item Encoding Time: averaged computing time per single 2-second raw seven-channel signal. The encoding times should be checked in relation to each other, rather than as absolute values, as they are recorded on the local machine.
    %\item AFR:  Average Firing Rate, defined as the number of spikes in a spike train over the total number of possible spike occurrences.
    %\item SNR: Signal to Noise Ratio, where the signal indicates the average power of the original signal, and the noise indicates the average power of the difference between the raw signal and the reconstructed signal.
    %\item Synaptic Operations: Sum of all spikes generated in the SNN, which can be used to indicate the energy consumption of the neural network model \cite{lemaire2022synaptic}.
    %\item Encoding Time: the time consumed to encode one instance to the event trains, measured on a local PC (xxxx).
    \item Accuracy: the classification accuracy of leaving one user out of cross-validation.
    \item Robustness: presented by the accuracy drop when introducing error spikes to the input spike trains.
    \item Deployment Performance: the energy and execution time when running an inference on Loihi2.
\end{itemize}

Negative spikes are introduced in the delta modulation where negative spikes naturally exist, and the TTFS with logarithmic distribution where positive spikes alone can not interpret the full range information. 
The binary encodings and multi-threshold delta modulation result in an extra dimension for the generated spike trains because of the parallel encoding and multiple thresholds, respectively.
Table \ref{tab:result_all} summarizes the evaluation result of each encoding scheme regarding the above-listed metrics.

%\subsection{Encoding Time}
%The binary and TTFS encoding schemes are the fastest methods to get the spike-form instances, consuming around 0.3-0.4 ms, indicating the low computational load of binary and TTFS encoding. Rate encoding schemes consume ten times more time using a probability mapping function and a Bernoulli trial. The multi-threshold delta modulation needs the longest time to get an event-form instance, as each data point (after the interpolation) needs to be compared with its predecessor, and the result needs to be compared again with five thresholds for spike generation. 

%\subsection{AFR}

AFR during encoding is an important metric indicating the spike density, which further influences the accuracy and robustness of the spiking neural network. For rate encoding, although different probability mapping functions are used to scale up the differences between numerical values of the special data set (most values are close to 0.5), nearly half of the time steps in an instance are fired, indicating the high input spike density compared to the other encoding schemes. The TTFS encoding has only one spike in one instance with 50 time steps, which is fragile to errors.

Another important aspect of spike encodings is how well they represent the original signal. The SNR is used to calculate the information loss of the various encodings. The spike rates in rate encoding come directly from the original numerical values or are fed through a random distribution first. For the uniform distribution, the rates are the corresponding approximation values. In the case of the normal and the beta distribution, the rates need to be given to their respective PPF to get the approximated values. For the TTFS encoding, decoding works by finding the index within the window where the spike is fired. In the case of the linear function, getting back to the original value is a linear calculation. The later the spike occurred, the smaller the value, or in the case of negative spikes, the further the value is away from 0.5. For the decoding with the logarithmic function, an exponential function needs to be constructed first. For the binary distribution, the spikes represent the binary fraction part of the index. The converted values from all trains are then added together to get the approximated value. As the 5-threshold is a relative encoding, only looking at the difference between two data points, decoding needs an initial value. The first value of the data signal is, therefore, read out and used to reconstruct the signal. As can be seen, the ten bits binary encoding gives the best SNR as a result of using the best resolution in signal encoding, while the linear TTFS falls behind in signal reconstruction. Generally, the rate encoding schemes using the normal and beta mapping function surpass most other schemes, which positively affects classification accuracy.

Regarding accuracy, most of the encoding schemes could supply an accuracy of around 0.9 after fine-tuning the hyperparameters using a grid-search way. The rate encoding with combined beta distribution supplies the best testing accuracy, while the rate encoding with uniform value-to-probability mapping function and the binary encoding with 6-bit resolution slightly fall behind. Surprisingly, although the TTFS with linear value-to-time mapping gives the lowest SNR, the classification accuracy reaches up to 89.1\%. This is explainable as the accuracy depends not only on the information loss during transformation but, most importantly, on the spike plasticity differences of the transformed spike-form instances.

The robustness is tested by introducing error spikes to the spike trains of each encoding scheme. Each spike has a chance of being changed, and the error probability indicates the probability a spike would be changed, e.g., vanishing or polarity exchange. For example, in a one-dimensional spike train with only 100 spike possibilities, an error probability of 0.1 would mean that with 10 indices, a spike would change from 1 to 0 or the other way around, respectively. When giving different error probabilities, the five-threshold delta modulation scheme shows the best robustness against others, while the TTFS is quite fragile, as it gives only one spike per data point. The binary encoding schemes surpass others besides the delta modulation when introducing larger errors. 

We finally deployed the spiking neural networks with different input spike trains to the Loihi 2 platform and recorded the averaged dynamic energy on the cores and the execution time of one instance. Generally, the energy consumed greatly depends on the time steps assigned to each spiking neural network, and the execution time remains close to each other, as the network model for each scheme is the same. Interestingly, such profiling, reported directly from the framework, showed a few negative dynamic energy numbers in some test cases, which could be a result of low-accuracy measurement of total and static energy. We also noticed that for each run, the energy and execution time vary a lot, which is described in other work \cite{shrestha2024efficient}.  Here, we simply reported the averaged number over a mass of on-device test.

\section{Conclusion and Future Work}

To use the computing efficiency of spiking neural networks in ubiquitous computing, especially where systems are battery-powered and real-time feedback is expected, e.g., haptic interfaces, this work evaluates different encoding schemes for sensor signals in a straightforward way, using a challenging public IMU-based dataset where the signal amplitude is unevenly distributed. 
Our evaluations show that each encoding schemes have its advantages and limitations, e.g. the Time-To-First-Spike encoding is easy to implement and needs the lowest firing rate during encoding while being able to achieve comparative classification accuracy; however, it is the worst scheme in resisting noisy spikes; The binary encoding could supply the best signal-to-noise ratio with higher resolution while giving relative low classification accuracy when encoding the signal with similar resolution; The multi-threshold delta modulation shows the best robustness and the rate encoding gives the best classification accuracy when using a proper value-to-probability mapping function. For future ubiquitous neuromorphic exploration, researchers should select the best encoding scheme depending on, first, the signal features like distribution status, second, the specific performance requirement, like robustness, accuracy, etc.

%\section*{Acknowledgment}

\bibliographystyle{IEEEtran}
\bibliography{reference}{}

\end{document}